\def\OO{\char'037}

\def\gapprox{\lower.4ex\hbox{$\;\buildrel >\over{\scriptstyle\sim}\;$}}
\def\lapprox{\lower.4ex\hbox{$\;\buildrel <\over{\scriptstyle\sim}\;$}}

\def\ref#1   {\par\noindent\hangindent1cm {#1}}
 \documentstyle[aaspp4]{article}        
 \input epsf
\slugcomment{{\sl }   \hfill{Revised Manuscript: 2003 Sept 11}}

\begin{document}

\title{         MHD Sausage Mode Oscillations in Coronal Loops		}
  
\author{	Markus J. Aschwanden\footnote{
		Lockheed Martin Advanced Technology Center,
                Solar \& Astrophysics Laboratory,
                Dept. L9-41, Bldg.252, 3251 Hanover St.,
                Palo Alto, CA 94304, USA;
                e-mail: aschwanden@lmsal.com},
		Valery M. Nakariakov\footnote{
		Physics Department, University of Warwick,
		Conventry, CV4 7AL, United Kingdom;
		valery@astro.warwick.ac.uk}, and
		Victor F. Melnikov\footnote{
		Radiophysical Research Institute (NIRFI),
		25 Bol'shaya Pecherskaya,
		Nizhny Novgorod 603950, Russia}}
		
\begin{abstract}
A recent study by Nakariakov et al. pointed out that the dispersion relation 
of MHD sausage mode oscillations has been incorrectly applied to coronal loops, 
neglecting the highly dispersive nature of the phase speed and the long-wavelength 
cutoff of the wave number. In the light of these new insights we revisit
previous observations that have been interpreted in terms of MHD sausage
mode oscillations in coronal loops and come to the following conclusions:
(1) Fast sausage MHD mode oscillations require such a high electron density
imposed by the wave number cutoff that they can only occur in flare loops;
(2) In the previously reported radio observations ($\nu \approx 100$ MHz to 1 GHz) 
with periods of $P\approx 0.5-5$ s, the fast sausage MHD mode oscillation is likely
to be confined to a small segment (corresponding to a high harmonic node) 
near the apex of the loop, rather than involving a global oscillation over 
the entire loop length. The recent microwave and soft X-ray observations 
of fast periods ($P\approx 6-17$ s) by Asai et al. and Melnikov et al., however,
are consistent with fast sausage MHD oscillations at the fundamental harmonic.
\end{abstract}


\section{              	INTRODUCTION 		}

Coronal seismology became an efficient new tool that uses standing MHD waves and 
oscillations as a tool to explore the physical parameters of the solar corona.
There are three basic branches of solutions of the dispersion relation for 
propagating and standing MHD waves: the slow mode branch (with acoustic
phase speeds), the fast mode branch and the Alfv\'en branch (with Alfv\'enic phase 
speeds). Furthermore, each branch has a symmetric and asymmetric solution, termed
the sausage and kink mode (Roberts et al. 1984). All of these MHD oscillation modes
have been now detected with imaging observations in recent years: transverse 
fast kink-mode oscillations with TRACE (Aschwanden et al. 1999; Nakariakov et al. 1999); 
longitudinal (slow-magnetoacoustic) modes with SUMER (Wang et al. 2002; Ofman \& Wang
2002), and fast sausage-mode oscillations probably with the {\sl Nobeyama Radioheliograph}
(Asai et al. 2001; Melnikov, Reznikova, \& Shibasaki 2002). The latter type of fast
sausage-mode oscillations is the least established one, although numerous non-imaging
radio observations exist that have been interpreted earlier in terms of this mode. However,
a recent study by Nakariakov et al. (2003) pointed out that the dispersion relation
and oscillation period has been incorrectly applied to the data, because the
highly dispersive nature of the phase speed and the long-wavelength cutoff in the 
wave number has been ignored. Here we revisit the relevant earlier radio observations
in the light of these new insights and attempt to derive the correct physical parameters. 

\section{ 		THEORY 			}

The dispersion relation for magnetosonic waves in cylindrical magnetic
flux tubes has infinitely many types of long-wavelength solutions in the fast-mode branch ($n=0,1,2,3,...$),
with the lowest ones called sausage modes ($n=0$) and kink modes ($n=1$). The solutions are depicted in Fig.~1,
showing that the kink mode solution extends all the way to the long-wavelength
limit $ka \mapsto 0$,
while the sausage mode has a cutoff at a phase speed of v$_{ph}=$v$_{Ae}$,
which has no solution for small wave numbers $ka < k_c a$. 
The propagation cutoff for the sausage mode occurs at the cutoff wave number $k_c$ 
(Edwin \& Roberts 1983; Roberts et al. 1984),
\begin{equation}
	k = k_c = \left[ { (c_s^2 + v_A^2) (v_{Ae}^2 - c_T^2) \over
			   (v_{Ae}^2 - v_A^2) (v_{Ae}^2 - c_s^2) }\	
		\right]^{1/2} \left({j_{0,s} \over a} \right)\ , \quad s=1,2,3,...
\end{equation}
where $j_{0,s} = (2.40, 5.52, ...)$ are the zeros of the Bessel function $J_0$.
Here we restrict our attention to the genuine sausage mode with transverse mode number $s=1$, 
i.e. $j_{0,1}$. The cutoff frequency $\omega$ at the cutoff is $k_c v_{Ae} = {\omega}_c$.  
Under coronal conditions, the sound speed amounts typically to $c_s = 150
\times \sqrt{T_{MK}} \approx 150-260$ km s$^{-1}$ (with $T\approx 1-3$ MK), 
which is much smaller than the typical Alfv\'en speed ($v_A \approx 1000$ km s$^{-1}$) 
so we have $c_s \ll v_A$. In this case the tube speed,
\begin{equation}
	c_T={ c_s^2 v_A^2 \over (c_s^2 + v_A^2)}	\ ,
\end{equation}
is similar to the sound speed, $c_T \lapprox c_s$, and the expression for the cutoff
wave number $k_c$ (Eq.~1) simplifies to, 
\begin{equation}
	k_c \approx \left({j_{0,s} \over a} \right) 
	\left[ { 1 \over (v_{Ae}/v_A)^2 - 1} \right]^{1/2}	
\end{equation}
In the low-$\beta$ corona, the thermal pressure is much smaller than the magnetic
pressure, so that we can assume almost identical magnetic field strengths
inside and outside of coronal loops, i.e., $B_0 \approx B_e$, so that
the ratio of external to the internal Alfv\'en speed $v_{Ae}/v_A$ is essentially
a density ratio $\sqrt{n_0/n_e}$,
\begin{equation}
	k_c \approx \left({j_{0,s} \over a} \right) 
	\left[ { 1 \over {n_0/n_e} - 1} \right]^{1/2}	
\end{equation}

From this expression we see that for typical density ratios inferred in the
solar corona, e.g., $n_e/n_0 \approx 0.1 - 0.5$ (Aschwanden et al. 2003), 
the cutoff wave number $k_c a$ falls into the range of $0.8 \lapprox k_c a \lapprox 2.4$. 
Therefore, we would expect that a long-wavelength 
sausage mode oscillation is completely suppressed for the slender loops
for which kink mode oscillations have been observed, which have wave numbers
of $k a = 2 \pi a / \lambda = (\pi /2 ) (w/l) \approx 0.04-0.08$ (Aschwanden et al. 2002).
Coronal conditions with $ka \ll 1$ were also inferred from fast kink mode oscillations
(Nakariakov \& Ofman 2001).
Of course, higher harmonics would have correspondingly shorter wavelengths, 
i.e. $k = N \pi/ l$ with $N=1,2,3,...$).
On the other side, the short-wavelength limit is given by the maximum possible loop width 
($a=w/2 \le l$), so the maximum wave number is 
$k a \leq k_m a = 2\pi (a/\lambda)_m = \pi (a/l)_m = \pi$.
The occurrence of global sausage mode oscillations therefore requires special
conditions, $k_c < k < k_m$, i.e., very high density contrast $n_0/n_e$ and relatively thick  
loops to satisfy $k > k_c$. The high density ratio, i.e., $n_0/n_e \gg 1$ or $v_{Ae} \gg v_A$, 
yields the following simple expressions for the cutoff wave number $k_c$,  
\begin{equation}
	k_c a \approx j_{0,s} \left( { v_A \over v_{Ae}} \right) 
	            = j_{0,s} \left( {n_e \over n_0} \right)^{1/2}	
\end{equation}
Since the wavelength of the fundamental eigenmode (with harmonic number $N=1$
in longitudinal direction and mode number $s=1$ in transverse direction)
corresponds to
the double loop length, so that the wavenumber relates to the loop length
by $k = 2\pi/\lambda = \pi/l$, the cutoff wave number condition $k > k_c$ implies a 
constraint between the loop geometry ratio $w/l$ and the density contrast
$n_e/n_0$,
\begin{equation}
	{ l \over w} = {\pi \over 2 a k} < {\pi \over 2 a k_c} =
	{\pi \over 2 j_{0,s}} \sqrt{ {n_0 \over n_e} } 
	\approx 0.65  \sqrt{ {n_0 \over n_e} } \ ,
\end{equation}
The numerical factor $0.65$ applies to the fundamental ($s=1$) sausage mode.
Since geometric parameters (such as $l$ and $w$) can be measured easier than
densities, we might turn the cutoff condition around and formulate it as a  
density contrast requirement for a given loop aspect ratio,
\begin{equation}
	{n_0 \over n_e} > \left( {1 \over 0.65} {l \over w} \right)^2
			= 2.4 \left( {l \over w} \right)^2 \ .
\end{equation}

This clearly indicates that slender loops with a high length-to-width ratio
$l/w \gg 1$ would require extremely high density contrast $n_0/n_e$.
Typical active region loops, which have only a moderate density contrast
in the order of $n_0/n_e \approx 2 - 10$, would be required to be 
extremely fat, with width-to-length ratios of $l/w \approx 1-2$. 
The density contrast is much higher for flare loops or postflare loops,
up to $n_0/n_e \approx 10^2, ... 10^3$. In this case, a length-to-width
ratio of $l/w \approx 6-20$ would be allowed for sausage mode oscillations.
This brings us to the conclusion that global sausage type oscillations are
only expected in fat and dense loops, basically only in flare and postflare
loops, a prediction that was not appreciated until recently (Nakariakov
et al. 2003). The restriction of the sausage mode wave number cutoff is
visualized in Fig.~2, where the permitted range of geometric loop aspect
ratios $(l/w)$ is shown as function of the density contrast $(n_0/n_e)$. 

\medskip
Since the sausage mode is highly dispersive 
in the long-wavelength part of the spectrum (see Fig.~1), the phase speed
$v_{ph}$ is a strong function of the wavenumber $ka$. The phase speed diagram
Fig.~1 shows that the phase speed equals to the external Alfv\'en speed
at the long-wavelength cutoff, $v_{ph}(k=k_c) = v_{Ae}$, and tends to approach
the internal Alfv\'en speed in the short-wavelength limit, $v_{ph}(ka \gg 1) \mapsto v_A$.
For coronal conditions, the phase speed of both the fast kink and fast sausage mode 
are bound by the internal and external Alfv\'en velocities,
\begin{equation}
	v_A \le v_{ph} = {\omega \over k} \le v_{Ae} \ ,
\end{equation}
Therefore, also the period ($P=2 l /v_{ph}$) of the standing sausage mode is 
bound by these two limits,
\begin{equation}
	{2 l \over v_{Ae}} < P_{saus} = {2 l \over v_{ph}} < {2 l \over v_A} \ .
\end{equation}
At the lower limit we can derive a simple relation for the sausage mode period
from the long-wavelength cutoff, where $v_{ph}(k=k_c)=v_{Ae}$, using Eq.~6,
\begin{equation}
	P_{saus} = {2 l \over v_{ph}} = {2 \pi \over k v_{ph}(k)}
	< {2 \pi \over k_c v_{ph}(k_c)} \approx 
	{2 \pi a \over j_{0,s} v_{Ae} } \left( {v_{Ae} \over v_A} \right) = 
	{2 \pi a \over j_{0,s} v_{A} } = {2.62 \ a \over v_A} \ .
\end{equation}
an approximation that is also derived in Rosenberg (1970) and Roberts et al. (1984). 
However, one should be aware that this relation represents only a lower limit
that applies at $k=k_c$, while for all other valid wavenumbers ($k > k_c$) the
loop length is shorter ($l=\pi/k$) and thus the global sausage mode period $P_{saus}$
is also shorter (Nakariakov et al. 2003), given by the actual phase speed $v_{ph}(k=\pi/l)$
defined by the dispersion relation (Fig.~1) at a particular value $k>k_c$, 
\begin{equation}
	P_{saus}(l) = {2 l \over v_{ph}(k=\pi/l)} \ ,
\end{equation}
Provided that sufficiently fat and overdense loops exist according to the requirement 
(Eq.~6 or 7), we expect for a loops radius of $a \approx 1000$ km and an 
Alfv\'en speed of $v_A \approx 1000$ km s$^{-1}$ a typical sausage mode period
of $P_{saus} \le 2.6$ s (according to Eq.~10).

\section{		OBSERVATIONS 		}

Table 1 contains a compilation of radio observations of oscillation events with 
periods in the range of 0.5 s $<P<$ 1 min. Most of these observations have been
interpreted in terms of MHD fast sausage mode oscillations, mainly for three
reasons: (1) Eigen-modes of standing MHD waves provide a natural mechanism 
for oscillations with a strictly regular period; (2) the period range of $P\approx 0.5-5$ s
was expected for the MHD fast sausage mode under coronal conditions, 
while other MHD oscillations modes have
significantly longer periods; and (3) the MHD fast sausage mode modulates the
loop density, magnetic field strength, and related radio emission 
(plasma emission or gyrosynchrotron emission), while the MHD kink mode does not
modulate the loop density in first order. In addition we might add a forth
reason which has not been appreciated previously: These events listed in Table 1
were generally reported to occur during flares, which explains the high densities 
that are expected on theoretical grounds (due to the sausage mode wave number cutoff 
criterion $k_c > k$, see Eq.~7), which does not occur for the MHD kink mode. 

Because the loop density plays a critical role for MHD fast sausage mode oscillations
(Eq.~7), we listed in Table 1 only observations that have information on the period $P$
as well on the electron density $n_0$ of the oscillating loops. 
There are only two observations with direct
electron density measurements, using soft X-ray emission measures and constraints
from the microwave spectrum, which report
electron densities of $n_0=4.5 \times 10^{10}$ cm$^{-3}$ (Asai et al. 2001) and
$n_0=10^{11}$ cm$^{-3}$ (Melnikov et al. 2002). However, we might also extract
information on the electron density from any radio burst that is produced by an
emission mechanism near the plasma frequency, since the plasma frequency depends
only on the electron density $n_0$,
\begin{equation}
	{\nu }_p = 8980 \sqrt{n_0 [{\rm cm}^{-3}]} \qquad {\rm [Hz]} \ . 
\end{equation}
Plasma emission is the dominant emission mechanism for metric and decimetric
radio bursts up to frequencies of $\nu \lapprox 1.5$ GHz, while free-free emission,
gyroresonance, and gyrosynchrotron emission dominate at higher frequencies (e.g., 
see Dulk 1985). A cursory look at Table 1 shows that most of the reported
oscillations fall in the frequency range of $\nu \approx 100-1000$ MHz, so this
corresponds to electron densities of $n_e \approx 10^8-10^{10}$ cm$^{-3}$ in the
case of plasma emission. In Fig.~3 we display the observed oscillation periods $P$
as function of the observed radio frequency ${\nu}$.  

\medskip
The observed radio frequency uniquely constrains the density $n_0$ of the oscillating
loop segment, for the case of plasma emission. In order to estimate the external
density $n_e$ we can use a standard model of the average background corona, for
instance the Baumbach-Allen model, which as function of the normalized height 
$R=(1+h/R_{\odot})$ is,
\begin{equation}
	n_e(h) = 10^8 \left[ {2.99 \over R^{16}} + {1.55 \over R^6} + {0.036 \over R^{1.5}} \right]
		\approx { 4\times 10^8 \over R^9 } \quad [{\rm cm}^{-3}]
\end{equation}
where the approximation on the right-hand-side holds for $h \lapprox R_{\odot}/2$.
This density model $n_e(h)$ is shown in Fig.~4 (top). Assuming that the fattest
possible loops have a width-to-length ratio of $q_w \approx 1/4$, the minimum
required density ratio is $n_{0,min}(h)/n_e(h) = 2.4/q_w^2 \approx 40$ (Eq.~7)
for loops oscillating in the MHD fast sausage mode due to the long-wavelength
cutoff criterion (Eq.~5). This cutoff criterion implies the constraint that
physical solutions are only possible for heights $h \gapprox 40$ Mm,  
for oscillations observed in the plasma frequency range of ${\nu}_p \lapprox 1$ 
GHz (grey area in Fig.4 top). The two events observed by Asai et al.
(2001) and Melnikov et al. (2002) do not fall in this range, because their
density is higher and would correspond to plasma
frequencies of ${\nu}_p=2$ and 3 GHz. The important conclusion that follows, based on the
two assumptions that the Baumbach-Allen model represents a realistic density
model of the background corona and that oscillating loops are not fatter than
a quarter of their length, is that physical solutions for the MHD fast sausage
mode are only possible for coronal heights of $h \gapprox 40$ Mm in 
the observed frequency range of ${\nu} \lapprox 1$ GHz, where most of
the fast oscillations have been observed (Fig.~3).

\medskip
Next we investigate the constraints introduced by the pulse periods, which
are found in the range of $P_{saus} \approx 0.5-5.0$ s (Fig.~3). Denoting the
length of the loop segment that participates in the MHD fast sausage mode
oscillation with ${\lambda}/2$, we obtain an upper limit from the phase speed
at the wavenumber cutoff $k_c$ (Eqs.~9, 11),
\begin{equation}
	{\lambda \over 2} = {P_{saus} \over 2} v_{ph}(k) \le {P_{saus} \over 2} v_{Ae} \ ,
\end{equation}
where the Alfv\'enic speed $v_{Ae}$ is given by the background density model $n_e(h)$ 
(Eq.~13),
\begin{equation}
	v_{Ae}(h) = {B \over \sqrt{ 4 \pi {\rho}_0 }} 
		  = {B \over \sqrt{ 4 \pi \mu m_H n_e(h) }} 
		  \approx 1210 \left( {B \over 20\ {\rm G}} \right)
			\left( {n_e(h) \over 10^9\ {\rm cm}^{-3}} \right)^{-1/2} \quad [{\rm km/s}]\ ,
\end{equation}
where we used a mean molecular weight of $\mu = 1.3$ for the H:He=10:1 coronal abundance.
We plot the sausage mode period $P_{saus}$ as function of the wavelength ${\lambda}/2$ in
Fig.~4 (bottom), for a mean coronal magnetic field of $B=20$ G (solid line), as well
as for a factor of 2 smaller or larger values ($B=10-40$ G). We find that the
corresponding wavelength of the oscillating loop segments have values in the range of 
$0.3$ Mm $< {\lambda}/2 <$ 10 Mm (gray area in Fig.~4 bottom) for pulse periods of
$0.5 < P_{saus} < 5.0$ s, which is much smaller than the full loop length 
($l=\pi h > 120$ Mm, see gray area in Fig.~4 top) required for the fundamental 
sausage mode. Because physical solutions are only possible in heights $h>40$ Mm, 
the oscillating loop segment has to be located near the top of the loop. Thus we 
conclude that only a small segment of the loop is oscillating in the MHD sausage mode. 
Such small oscillating loop segments require wavelengths that correspond to higher 
harmonics $N$ of the fast sausage MHD mode, 
\begin{equation}
	\lambda = {2 \pi \over k} = {2 l \over N} \ ,
\end{equation}
while lower harmonics cannot oscillate due to the wave number cutoff. 
The situation is illustrated for the case $N=3$ in Fig.~5:
For instance, if the cutoff is $k \ge k_c = 6 \pi /l$, only wavelengths with 
$\lambda = 2 \pi / k \le 2 \pi / k_c = l/3$ are able to oscillate, which is most 
likely to occur at the loop top where the density contrast and wavelength match 
the sausage wavenumber cutoff criterion. 
Alternatively, oscillations at a higher
harmonic could occur along the entire loop, but favorable conditions for
detection could be restricted to certain segments only (e.g., due to 
line-of-sight angles or absorption of plasma emission).

An alternative interpretation of these fast radio pulsations in the decimetric
range could be made in terms of propagating fast sausage mode MHD waves, where the
wavelength is not prescribed by the loop length, but independently by the driver. 
This interpretation has been proposed in Roberts et al.~(1984). However, if this
would be the case, the radio dynamic spectra should show parallel drifting bands
of emission for every propagating node. Such decimetric burst types have indeed
been observed and are called {\sl decimetric fiber bursts} (e.g., Rosenberg 1972;
Bernold 1980; Slottje 1981), but many of the observations listed in Table 1 show
narrowband pulsations without parallel drifting bands, and thus are not 
consistent with (impulsively generated) propagating fast sausage MHD waves. 

Note that the measurements of Asai et al. (2001) and Melnikov et al. (2002) have
longer periods and higher densities than all decimetric observations listed in Table 1, 
and thus actually allow for global sausage mode oscillations of the entire
loop length, for reasonable magnetic fields in the order of $B\approx 40$ G (marked
with crosses in Fig.~4 bottom).

\section{		CONCLUSIONS		}

This study leads us to the following conclusions:
\begin{enumerate}
\item{  Only the two imaging observations of fast oscillation events ($P \approx 10$ s)
	by Asai et al. (2001) and Melnikov et al. (2002) are consistent with
	global MHD fast sausage mode oscillations, satisfying the relation between
	period and loop lengths as well as the wave number cutoff criterion for
	the fundamental sausage mode. The required magnetic field is of the order
	of $B\approx 40$ G.}
\item{	All earlier radio observations of fast oscillation events, which have been reported
	in the frequency range of ${\nu}=$100 MHz $-$ 1 GHz and with periods of
	$P \approx 0.5-5.0$ s, imply for reasonable coronal magnetic field strengths
	($B \approx 10-40$ G) that the length of the loop segments participating
	in the MHD fast sausage mode oscillation is much smaller ($\lambda/2 \lapprox
	10$ Mm) than the full loop length ($l\approx 100-500$ Mm). The sausage mode
	oscillation seems to be confined to segments where the largest density contrast to 
	the background occurs, most likely near the loop top, in a segment corresponding
	to a higher harmonic node (see Fig.~5).}
\item{	The wave number cutoff criterion seems to introduce a bimodality of
	MHD fast mode sausage oscillation phenomena: (1) partial loop oscillations
	confined to loop segments ($\lambda/2 \ll l$) of high harmonic $(N \gg 1)$ nodes
	for low-density ($n_e \lapprox 10^{10}$ cm$^{-3}$) loops that are observable
	at the plasma frequency in the metric/decimetric frequency range 
	(${\nu} \lapprox 1$ GHz); and
	(2) global loop oscillations ($\lambda/2 = l$) at the fundamental mode ($N=1$)
	for high-density ($n_e \approx 10^{10}-10^{11}$ cm$^{-3}$) loops that are 
	observable in soft X-rays and via gyrosynchrotron emission in microwaves.}
\end{enumerate} 
In this study we explored the diagnostic of MHD fast sausage mode oscillations.
It is important to notice that the observed period does not provide a diagnostic of
the Alfv\'enic travel time across the loop diameter ($P = 2.62 a / v_A$), as it was 
commonly believed, but rather provides a lower limit of this quantity (Nakariakov et al. 2003). 
Moreover, the new finding that fast periods are likely to be caused by oscillations
of partial loop segments (at higher harmonic nodes), 
implies that the magnetic field $B$ cannot be determined 
by the loop diameter and sausage mode period, unless the harmonic number can be
constrained from imaging observations. This study illustrates specific 
constraints that are contingent on the sausage wave number cutoff, which have been
ignored in previous studies.  

\subsection*{Acknowledgements:}
MJA was partially supported by NASA contract NAS5-38099 (TRACE),
VFM was supported by RFBR grant No.02-02-39005 and NSF grant AST-0307670.

\subsection*{References} 

\newcommand{\AaA}{AA,\ }                
\newcommand{\AAR}{AAR,\ }               
\newcommand{\AAS}{AAS,\ }               
\newcommand{\ApJ}{ApJ,\ }               
\newcommand{\ApJL}{ApJL,\ }             
\newcommand{\ApJS}{ApJS,\ }             
\newcommand{\ARAA}{ARAA,\ }             
\newcommand{\EAA}{EAA,\ }               
\newcommand{\GRL}{GRL,\ }               
\newcommand{\JGR}{JGR,\ }               
\newcommand{\MNRAS}{MNRAS,\ }           
\newcommand{\PASJ}{PASJ,\ }             
\newcommand{\SoP}{Solar Phys.,\ }        

\ref{Abrami,A. 1970, \SoP 11, 104.}
\ref{Abrami,A. 1972, Nature 238/80, 25.}
\ref{Achong,A. 1974, \SoP 37, 477-482.}
\ref{Asai,A., Shimojo, M., Isobe, H., Morimoto, T., Yokoyama, T., Shibasaki, K., and Nakajima, H.
 	2001, ApJ 562, L103.}
\ref{Aschwanden,M.J. 1986, \SoP 104, 57-65.}
\ref{Aschwanden,M.J. \& Benz,A.O. 1986, \AaA 158, 102-112.}
\ref{Aschwanden,M.J. 1987b, {\sl Pulsations of the radio emission of the solar corona.
        Analysis of observations and theory of the pulsating electron-cyclotron maser},
        PhD Thesis, ETH Zurich, pp.1-173.}
\ref{Aschwanden,M.J., Bastian,T.S. \& Gary, D.E., 1992c, {\sl Bull.Amer.Astron.Soc.} 24/2, 802.}
\ref{Aschwanden,M.J., Benz,A.O., \& Montello,M. 1994a, \ApJ 431, 432-449.}
\ref{Aschwanden,M.J., Benz,A.O. Dennis,B.R. \& Kundu,M.R. 1994b, \ApJS 90, 631-636.}
\ref{Aschwanden,M.J., Fletcher,L., Schrijver,C., and Alexander,D.,
 	1999, \ApJ, 520, 880.}
\ref{Aschwanden,M.J., DePontieu,B., Schrijver,C.J., and Title,A.
 	2002, \SoP 206, 99.}
\ref{Aschwanden,M.J., Nightingale,R.W., Andries,J., Goossens,M., and Van Doorsselaere,T.
 	2003, \ApJ subm.}
\ref{Aurass,H., Chernov,G.P., Karlicky,M., Kurths,J., \& Mann,G. 1987, \SoP 112, 347-357.}
\ref{Bernold,T.E.X. 1980, AAS 42, 43-58.}
\ref{Chernov,G.P. \& Kurths,J 1990, Sov. Astron. 34(5), 516-522.}
\ref{Chernov,G.P., Markeev,A.K., Poquerusse,M., Bougeret,J.L., Klein,K.L. Mann,G.,
        Aurass,H., \& Aschwanden,M.J. 1998, \AaA 334, 314-324.}
\ref{DeGroot,T. 1970, \SoP 14, 176.}
\ref{Dr{\"o}ge,F. 1967, Zeitschr.Astrophys. 66, 200.}
\ref{Dulk,G.A. 1985, Annu. Rev. Astron. Astrophys. 23, 169.}
\ref{Edwin,P.M. and Roberts,B. 1983, \SoP 88, 179.}
\ref{Elgaroy,\OO. 1980, \AaA 82, 308-313.}
\ref{Gotwols,B.L. 1972, \SoP 25, 232-236.}
\ref{Kai,K. \& Takayanagi,A. 1973, \SoP 29, 461-475.}
\ref{Kliem,B., Karlicky,M., \& Benz,A.O. 2000, \AaA 360, 715-728.}
\ref{Kurths,J. \& Herzel,H. 1986, \SoP 107, 39-45.}
\ref{Kurths,J. \& Karlicky,M. 1989, \SoP 119, 399-411.}
\ref{Kurths,J., Benz,A.O., \& Aschwanden,M.J. 1991, \AaA 248, 270-276.}
\ref{McLean,D.J, Sheridan,K.V., Steward,R.T., \& Wild,J.P. 1971, Nature 234, 140-142.}
\ref{McLean,D.J. \& Sheridan,K.V. 1973, \SoP 32, 485-489.}
\ref{Melnikov,V.F., Reznikova,V.E., \& Shibasaki,K. 2002, in Proc. of intern. Conf.
	{\sl Active Processes on the SUn and Stars}, (eds. Zaitsev,V.V. and Yasnov,L.V., p.225.}
\ref{Nakariakov,V.M., Ofman,L., DeLuca,E., Roberts,B., Davila,J.M.
 	1999, Science 285, 862.}
\ref{Nakariakov,V.M. \& Ofman,L. 2001, \AaA 372, L53.}
\ref{Nakariakov,V.M., Melnikov,V.F., and Reznikova,V.E. 2003, \AaA, subm.}
\ref{Ofman,L. and Wang,T.J. 2002, \ApJ 580, L85.}
\ref{Pick, M. \& Trottet, G. 1978, \SoP 60, 353-359.}
\ref{Roberts,B., Edwin,P.M., and Benz,A.O. 1984, \ApJ 279, 857.}
\ref{Rosenberg,H. 1970, \AaA 9, 159.}
\ref{Rosenberg,H. 1972, \SoP 25, 188-196.}
\ref{Sastry,Ch.V., Krishan,V., \& Subramanian,K.R. 1981, J. Astrophys. Astron. 2, 59-65.}
\ref{Slottje, C. 1981, {\sl Atlas of fine structures of dynamic spectra of
                solar type IV-dm and some type II radio bursts}, Dwingeloo, The Netherlands.}
\ref{Tapping,K.F. 1978, \SoP 59, 145-158.}
\ref{Trottet, G., Kerdraon, A., Benz, A.O., \& Treumann, R. 1981, \AaA 93, 129-135.}
\ref{Wang,T.J., Solanki,S.K., Curdt,W., Innes,D.E., and Dammasch,I.E.
 	2002, \ApJ 574, L101.}
\ref{Wiehl,H.J., Benz,A.O., \& Aschwanden,M.J. 1985, \SoP 95, 167-179.}
\ref{Zaitsev,V.V., Stepanov,A.V., \& Chernov,G.P. 1984, \SoP 93, 363-377.}
\ref{Zlobec, P., Messerotti, M., Dulk, G.A., \& Kucera, T. 1992, \SoP 141 165-180.}

\clearpage

\begin{deluxetable}{lllll}
\tiny
\tablecaption{Observations of oscillations in the period range of 0.5 s $<P<$ 1 min
with density information (either from radio frequencies of $\nu < 1.5$ GHz or
from soft X-rays emission measures).}

\tablehead{
\colhead{Observer}&
\colhead{Number}&
\colhead{Frequency}&
\colhead{Period}&
\colhead{Instrument/}\\
\colhead{}&
\colhead{of events}&
\colhead{$\nu$ [MHz]}&
\colhead{$P$ [s]}&
\colhead{Spatial resolution}}
\startdata

Dr{\"o}ge (1967)                & $>18$ & 240, 460 MHz  & 0.2-1.2       & Kiel\\
Rosenberg (1970)                & 1     & 220-320 MHz   & 1.0-3.0       & Utrecht\\
De Groot (1970)                 & $>25$ & 250-320 MHz   & 2.2-3.5       & Utrecht\\
Abrami (1970, 1972)             & 3     & 239 MHz       & 1.7-3.1       & Trieste\\
McLean et al. (1971)            & 1     & 100-200 MHz   & 2.5-2.7       & Culgoora\\
Rosenberg (1972)                & 1     & 220-320 MHz   & 0.7-0.8       & Utrecht\\
Gotwols (1972)                  & 1     & 565-1000 MHz  & 0.5           & Silver Spring\\
Kai \& Takayanagi (1973)        & 1     & 160 MHz       & $<$1.0        & 17' (Nobeyama)\\
McLean \& Sheridan (1973)       & 1     & 200-300 MHz   & 4.28$\pm$0.01 & Culgoora\\
Achong (1974)                   & 1     & 18-28 MHz     & 4.0           & Kingston\\
Tapping (1978)                  & 14    & 140 MHz       & 0.06-5        & Cranleigh\\
Pick \& Trottet (1978)          & 1     & 169 MHz       & 0.37, 1.7     & 5'-7' (Nan\c{c}ay)\\
Elgaroy (1980)                  & 8     & 310-340 MHz   & 1.1           & Oslo\\
Bernold (1980)                  & $>13$ & 100-1000 MHz  & 0.5-5         & Zurich\\
Slottje (1981)                  & $>40$ & 160-320 MHz   & 0.2-5.5       & Dwingeloo\\
Trottet et al. (1981)           & 1     & 140-259 MHz   & 1.7$\pm$0.5   & 5' (Nan\c{c}ay)\\
Sastry et al. (1981)            & 1     & 25-35 MHz     & 2-5           & 30' (Gauribidanur)\\
Zlobec et al. (1992)            & 1     & 333 MHz       & 9.8-14.2      & 0.7'-1.5' (VLA) \\
Zaitsev et al. (1984)           & 23    & 45-230 MHz    & 0.3-5         & Izmiran\\
Wiehl et al. (1985)             & 1     & 300-1000 MHz  & 1-2           & Zurich\\
Aschwanden (1986, 1987b)        & 10,60 & 300-1100 MHz  & 0.4-1.4       & Zurich\\
Aschwanden \& Benz (1986)       & 10    & 237, 650 MHz  & 0.5-1.5       & Zurich\\
Kurths \& Herzel (1986)         & 1     & 480-800 MHz   & 1.0, 3.5      & Tremsdorf\\
Aurass et al. (1987)            & 1     & 234 MHz       & 0.25-2        & Tremsdorf\\
Kurths \& Karlicky (1989)       & 1     & 234 MHz       & 1.3, 1.5      & Tremsdorf\\
Chernov \& Kurths (1990)        & 10    & 224-245 MHz   & 0.35-1.3      & Izmiran\\
Kurths et al. (1991)            & 25    & 234-914 MHz   & 0.07-5.0      & Zurich\\
Aschwanden et al. (1992c)       & 1     & 1.5 GHz       & 8.8           & 0.2'-0.9' (VLA, OVRO)\\
Aschwanden et al. (1994a,b)     & 1,1   & 300-650 MHz   & 1.15, 1.8     & Zurich\\
Chernov et al. (1998)           & 1     & 164-407 MHz   & 0.2           & 5'-7'(Nan\c{c}ay)\\
Kliem et al. (2000)             & 1     & 600-2000 MHz  & 0.5-3.0       & Zurich\\
Asai et al. (2001)              & 1     & 17 GHz        & 6.6           & 10" (Nobeyama)\tablenotemark{a}\\
Melnikov et al. (2002)          & 1     & 17, 34 GHz    & 8-11, 14-17   & 10",5" (Nobeyama)\tablenotemark{b}\\
\enddata
\tablenotetext{a}{Asai et al. (2001) estimated a loop length of $L=16$ Mm, a loop width of $w=6$ Mm,
	and a loop density of $n_e=4.5 \times 10^{10}$ cm$^{-3}$, using {\sl Nobeyama} and
	{\sl Yohkoh/SXT} data observed during the 1998-Nov-10, 00:12 UT, flare.}
\tablenotetext{b}{Melnikov et al. (2002) estimated a loop length of $L=25$ Mm, a loop width
	of $w=6$ Mm, and a loop density of $n_e=10^{11}$ cm$^{-3}$, using {\sl Nobeyama} and
	{\sl Yohkoh/SXT} observed during the 2000-Jan-12, 01:35 UT, flare.}
\end{deluxetable}


\begin{figure} 
\plotone{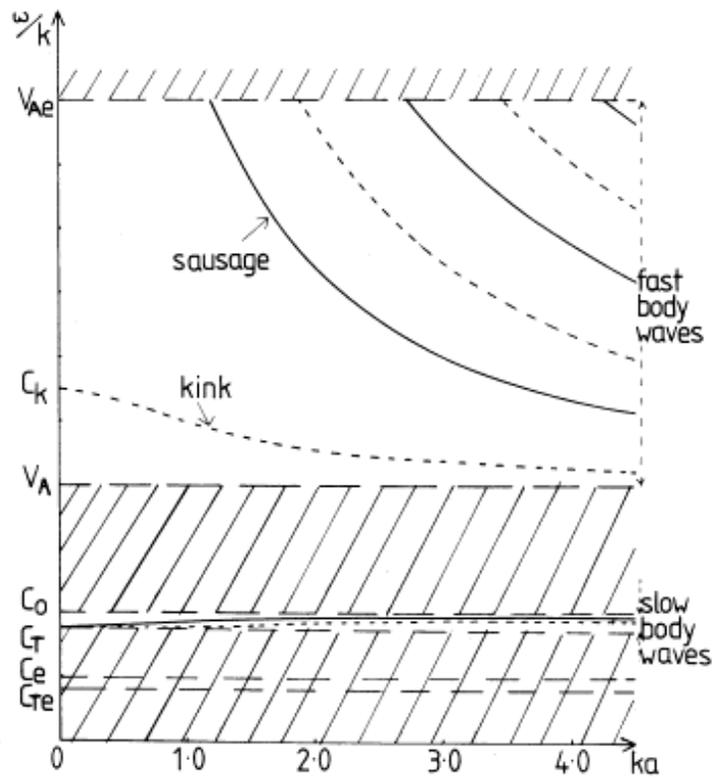}
\caption{The phase speed $\omega / k$ is shown for magnetoacoustic waves in
a cylindrical fluxtube (with radius $a$), as function of the longitudinal
wave number $ka$, for coronal conditions $v_{Ae} > v_A > c_t > c_s$. The
sausage modes are indicated with solid lines, the kink modes with dashed
lines (Edwin \& Roberts 1983).}
\end{figure}

\begin{figure} 
\plotone{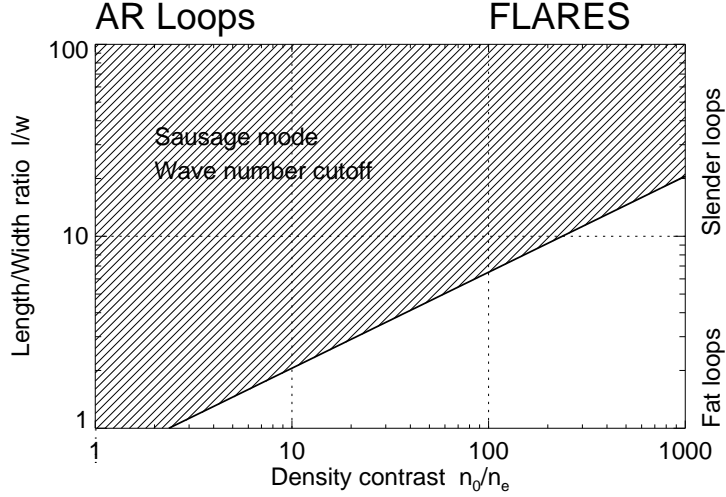}
\caption{The wave number cutoff $k_c$ for sausage mode oscillations
expressed as a requirement of the loop length-to-width ratio $l/w$
as function of the density contrast $n_0/n_e$ between the external
and internal loop densities (see Eq.~6). Note that sausage
mode oscillations in slender loops can occur only for flare conditions.}
\end{figure}

\begin{figure} 
\plotone{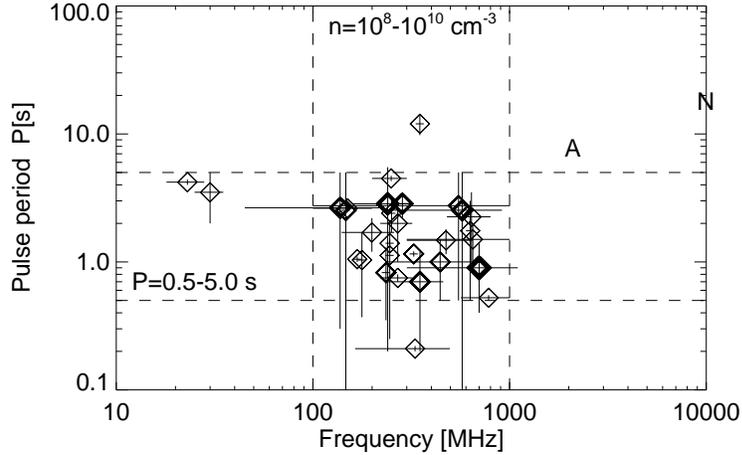}
\caption{Observed oscillation periods as function of the radio frequency,
according to the list given in Table 1. Horizontal and vertical bars indicate
the ranges in each observation, diamonds the mean values, and thick diamonds
label reports on multiple oscillation events. Note that most of the events
fall in the range of $P\approx 0.5-5.0$ s and $n_0=10^8-10^{10}$ cm$^{-3}$,
if one assumes that the radio frequency is near the plasma frequency, $\nu \approx {\nu}_p$.
The letter A and M refer to the observations of Asai et al. (2001) and Melnikov et al. (2003).}
\end{figure}

\begin{figure} 
\plotone{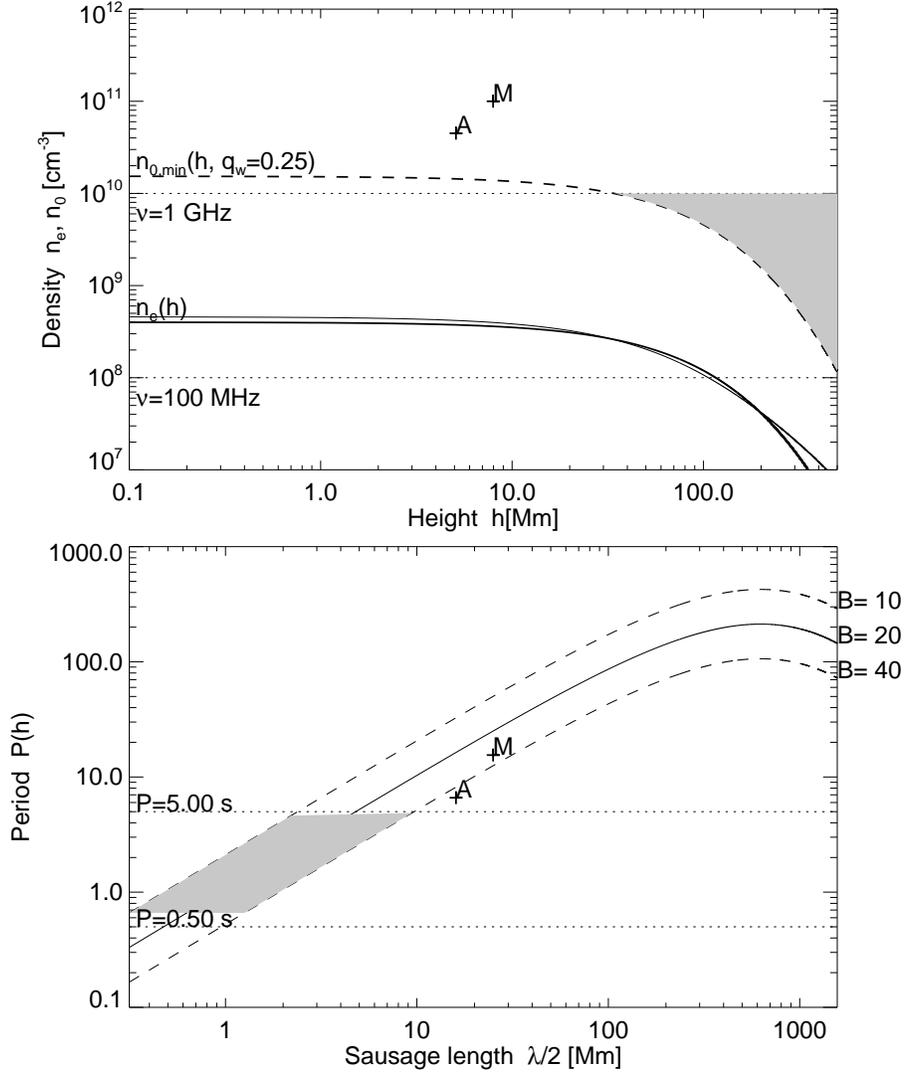}
\caption{{\sl Top:} The Baumbach-Allen density is shown as a model of the average coronal 
background (thin curve), together with the approximation $n_e(h) \approx 4\times 10^8
 (1 + h/R_{\odot})^9$ (thick curve), and the minimum loop density 
$n_{0,min}(h)$ required 
for loops that have a width-to-length ratio of $q_w=w/l \le 0.25$. The height and density
range corresponding to a plasma frequency range of 100 MHz-1 GHz is indicated with a
grey area. {\sl Bottom:} The MHD fast sausage mode period $P_{saus}$ is shown as
function of the sausage length ${\lambda}/2$ for magnetic fields of $B=20$ G within a
factor of 2. The regime of physical solutions for periods $P=0.5-5$ s and $B=10-40$ G
is indicated with a grey area. A and M indicate the values inferred for the observations 
of Asai et al. (2001) and Melnikov et al. (2002).}
\end{figure}

\begin{figure} 
\plotone{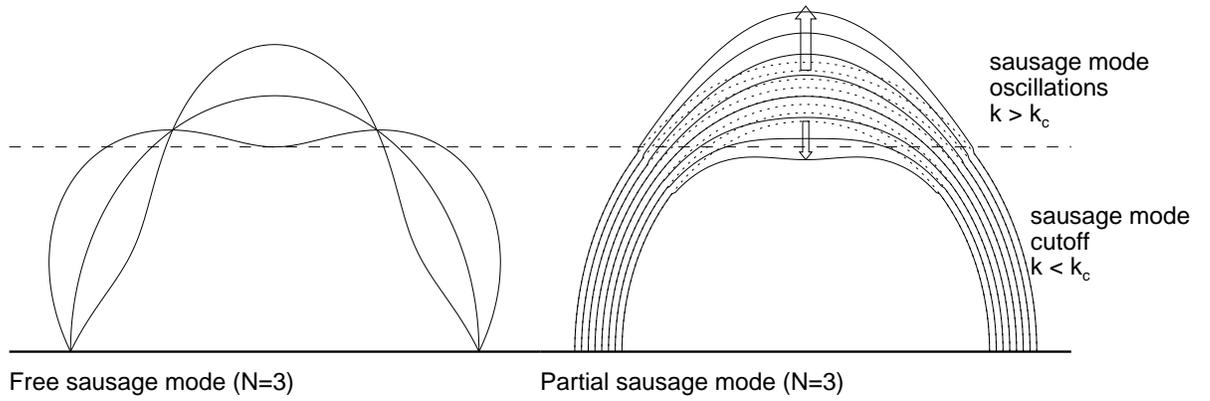}
\caption{Sketch of MHD fast sausage mode oscillation in partial loop segment (right), 
where the cutoff condition $k>k_c$ is satisfied. For a given width-to-length ratio $q_w$,
oscillations occur in segments where $n_0/n_e > 2.4/q_w^2$ (Eq.~7). The oscillating
segment may correspond to a higher harmonic node, e.g., to $N=3$ here (left).}
\end{figure}

\end{document}